\documentclass[iop,apj]{emulateapj}
\usepackage{graphics}
\usepackage{graphicx}
\usepackage{natbib}
\usepackage{epsfig}

\usepackage{epstopdf}
\begin{document}

\title{Modeling Multi-Wavelength Stellar Astrometry. III. Determination of the Absolute Masses of Exoplanets and Their Host Stars}
\author{Jeffrey L. Coughlin\altaffilmark{1,4} and Mercedes L\'opez-Morales\altaffilmark{2,3}}
\journalinfo{Accepted for Publication in the Astrophysical Journal}

\altaffiltext{1}{Department of Astronomy, New Mexico State University, P.O. Box 30001, MSC 4500, Las Cruces, New Mexico 88003-8001}
\altaffiltext{2}{Institut de Ci\`encies de l'Espai (CSIC-IEEC), Campus UAB, Facultat de Ci\`encies, Torre C5, parell, 2a pl, E-08193 Bellaterra, Barcelona, Spain}
\altaffiltext{3}{Carnegie Institution of Washington, Department of Terrestrial Magnetism, 5241 Broad Branch Road NW, Washington, DC 20015-1305, USA; Visiting Investigator}
\altaffiltext{4}{NSF Graduate Research Fellow}
\email{jlcough@nmsu.edu}

\begin{abstract}
Astrometric measurements of stellar systems are becoming significantly more precise and common, with many ground and space-based instruments and missions approaching 1 $\mu$as precision. We examine the multi-wavelength astrometric orbits of exoplanetary systems via both analytical formulae and numerical modeling. Exoplanets have a combination of reflected and thermally emitted light that cause the photocenter of the system to shift increasingly farther away from the host star with increasing wavelength. We find that, if observed at long enough wavelengths, the planet can dominate the astrometric motion of the system, and thus it is possible to directly measure the orbits of both the planet and star, and thus directly determine the physical masses of the star and planet, using multi-wavelength astrometry. In general, this technique works best for, though is certainly not limited to, systems that have large, high-mass stars and large, low-mass planets, which is a unique parameter space not covered by other exoplanet characterization techniques. Exoplanets that happen to transit their host star present unique cases where the physical radii of the planet and star can be directly determined via astrometry alone. Planetary albedos and day-night contrast ratios may also be probed via this technique due to the unique signature they impart on the observed astrometric orbits. We develop a tool to examine the prospects for near-term detection of this effect, and give examples of some exoplanets that appear to be good targets for detection in the K to N infrared observing bands, if the required precision can be achieved.
\end{abstract}


\section{Introduction}
\label{intro}

As part of a Space Interferometry Mission (SIM) Science Study, in \citet{Coughlin2010a}, hereafter referred to as Paper I, we examined the implications that multi-wavelength microarcsecond astrometry has for the detection and characterization of interacting binary systems. In Paper I we found that the astrometric orbits of binary systems can vary greatly with wavelength, as astrometric observations of a point source only measure the motion of the photocenter, or center of light, of the system. For systems that contain stellar components with different spectral energy distributions, the motion of the photocenter can be dominated by the motion of either component, depending on the wavelength of observation. Thus, with multi-wavelength astrometric observations it is possible to measure the individual orbit of each component, and thus derive absolute masses for both objects in the system. In \citet*{Coughlin2010b}, hereafter referred to as Paper II, we showed that multi-wavelength astrometry can also be used to directly measure the inclination and gravity darkening coefficient of single stars, as well as the temperature, size, and position of star spots.

Astrometry has long been used to measure fundamental quantities of binary stars, and more recently has been used to study extrasolar planets. Although no independently confirmed planet has yet been initially discovered via astrometry, many planets discovered via radial-velocity, (which only yields the planetary mass as a function of the system's inclination and host star's mass), have had follow-up astrometric measurements taken in order to determine their inclinations, and thus true planetary mass as a function of only the assumed stellar mass \citep{McArthur2004,Benedict2006,Bean2007,Martioli2010,McArthur2010,Roll2010,Reffert2011}. There are many ground and space-based microarcsecond precision astrometric projects which are either currently operating or on the horizon. The proposed SIM Lite Astrometric Observatory, a redesign of the earlier proposed SIM PlanetQuest Mission, was to be a space-based 6-meter baseline Michelson interferometer capable of 1 $\mu$as precision measurements in $\sim$80 spectral channels spanning 450 to 900 nm \citep{SIM2009}, thus allowing multi-wavelength microarcsecond astrometry. Although the SIM Lite mission has been indefinitely postponed at the time of this writing, it has already achieved all of its technological milestones, and it, or another similar mission, could be launched in the future. The PHASES project obtained as good as 34 $\mu$as astrometric precision of close stellar pairs \citep{Muterspaugh2010}. The CHARA array has multi-wavelength capabilities, and can provide angular resolution to $\sim$200 $\mu$as \citep{Brummelaar2005}. PRIMA/VLTI is working towards achieving $\sim$30-40 $\mu$as precision in the K-band \citep{Belle2008}, with GRAVITY/VLTI expected to obtain 10 $\mu$as \citep{Kudryavtseva2010}. The ASTRA/KECK project will be able to simultaneously observe and measure the distance between two objects to better than 100 $\mu$as precision. The GAIA mission will provide astrometry for $\sim$10$^{9}$ objects with 4 - 160 $\mu$as accuracy, for stars with V = 10-20 mag respectively, and does posses some multi-wavelength capabilities \citep{Cacciari2009}. The MICADO instrument on the proposed E-ELT 40-meter class telescope will be able to obtain better than 50 $\mu$as accuracy at 0.8-2.5 $\mu$m \citep{Trippe2010}. Finally, the NEAT mission proposes to obtain as low as 0.05 $\mu$as astrometric measurements at visible wavelengths \citep{Malbet2011}. Thus, astrometric measurements of extrasolar planets are going to become significantly more common in the future.

In this paper, we examine the multi-wavelength astrometric signature of exoplanets. A star-planet system is a specialized case of a binary system with extreme mass and temperature ratios, and thus the findings of Paper I apply to exoplanets. Specifically, an extrasolar planet has a combination of reflected and thermally emitted light that cause the photocenter to be displaced from the center of mass of the star. Since the planet's temperature is very different from that of the host star, the amount of photocenter displacement due to the planet will greatly vary with wavelength. Although the luminosity ratio between a star and planet is extreme, the planet also lies a much farther distance from the barycenter of the system compared to the star, and thus it has a large ``moment-arm'' with which to influence the photocenter. While conventional single-wavelength astrometric measurements can yield the inclination and spatial orientation of a system's orbital axis, with multi-wavelength astrometry it should be possible to measure the individual orbits of both the star and planet, and thus determine the absolute masses of both.

In \S\ref{eqsec} we derive analytical formulae for estimating the astrometric motion of a star-planet system at a given wavelength. In \S\ref{numsec} we perform numerical simulations of the multi-wavelength astrometric orbits of a few systems of interest using the {\sc reflux} code, and examine a few features specific to transiting planets. In both sections we present the most promising systems for future observation and detection of this effect. Finally, in \S\ref{discusssec} we discuss our results and what future work is needed to achieve these observations.

\section{Analytical Formulae for Computing the Reflex Motion}
\label{eqsec}

\begin{figure}
\centering
\epsfig{width=\linewidth,file=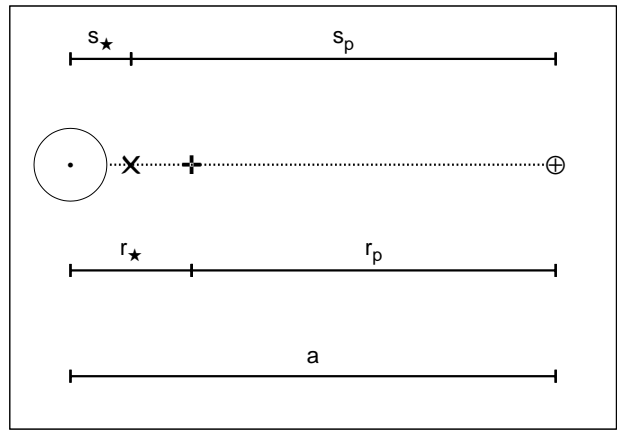}
\caption{An illustration of a system containing a star, shown on the left, and a planet, shown on the right, separated by a distance $a$, not to scale. The star and planet lie at distances of $r_{\star}$ and $r_{p}$, respectively, from the barycenter of the system, which is marked via a ``+'' symbol. Similarly, the star and planet lie at distances of $s_{\star}$ and $s_{p}$, respectively, from the photocenter of the system, which is marked via a ``$\times$'' symbol. All distances are sky-projected distances along the semi-major axis of the system, and thus are independent of the system's inclination. Note that although in this illustration the photocenter is to the left of the barycenter, it can lie anywhere between the star and planet.}
\label{schematicfig}
\end{figure}

Our objective is to derive an analytical expression for the amplitude of the sky-projected angular astrometric reflex motion of a star-planet system with respect to the wavelength of observation, $\alpha$. In all of the following equations, we are dealing with sky-projected distances measured along the semi-major axis of the system, and thus they are independent of the inclination of the system. We consider the case of a star and single planet in a circular orbit, with masses $M_{\star}$ and $M_{p}$ respectively, separated by an orbital distance, $a$, as illustrated in Figure~\ref{schematicfig}. The system's barycenter, marked via a ``+'' symbol, lies in-between the star and planet, at a distance of $r_{\star}$ from the star, and $r_{p}$ from the planet.

Defining the mass ratio, $q$, as

\begin{equation}
 q = \frac{M_{p}}{M_{\star}} 
\end{equation}

\noindent the values for $r_{\star}$ and $r_{p}$ are then

\begin{equation}
r_{\star} = \frac{a \cdot q}{q+1}
\end{equation}

\begin{equation}
 r_{p} = \frac{a}{q + 1}
\end{equation}

\noindent where by definition $r_{\star}$ + $r_{p}$ = $a$. 

In the case where all the light from the system is assumed to come from the star, i.e., the system's photocenter is the star's center, the wavelength-independent amplitude of the angular astrometric reflex motion of the system, $\alpha_{0}$, is

\begin{equation}
\label{alphaeq1}
\alpha_{0} = \arctan \left(\frac{r_{\star}}{D}\right) = \arctan\left(\frac{a \cdot q}{D\cdot(q+1)}\right)
\end{equation}

\noindent where $D$ is the distance to the system from Earth, and $a$, via Kepler's third law, is

\begin{equation}
\label{aeq}
 a = \left(G(M_{\star} + M_{p})\right)^{\frac{1}{3}}\left(\frac{P}{2\pi}\right)^{\frac{2}{3}}
\end{equation}

\noindent where $G$ is the gravitational constant, and $P$ is the orbital period of the system.

When the planet's luminosity is not negligible, in order to determine the wavelength-dependent value of $\alpha$, the location of the system's photocenter, which varies with wavelength, must be determined. We define $s_{\star}$ and $s_{p}$ to be the distance to the system's photocenter from the star and planet respectively, as shown in Figure~\ref{schematicfig}, where the photocenter is marked with a ``$\times$'' symbol. We define the luminosity ratio at a given wavelength, $L_{r}$, as

\begin{equation}
 L_{r} = \frac{L_{p}}{L_{\star}}
\end{equation}

\noindent where $L_{p}$ is the luminosity of the planet, and $L_{\star}$ is the luminosity of the star. Thus, similar to the previously presented derivations, the values for $s_{\star}$ and $s_{p}$ are

\begin{equation}
 s_{\star} = \frac{a \cdot L_{r}}{L_{r}+1}
\end{equation}

\begin{equation}
 s_{p} = \frac{a}{L_{r} + 1}
\end{equation}

\noindent where by definition $s_{\star}$ + $s_{p}$ = $a$. The observed astrometric motion results from the movement of the system's photocenter around the system's barycenter. Thus, taking into account light from both the star and planet,

\begin{equation}
 \alpha = \arctan\left(\frac{r_{\star} - s_{\star}}{D}\right) = \arctan\left(\frac{s_{p} - r_{p}}{D}\right) 
\end{equation}

\noindent and thus

\begin{equation}
\label{alphaeq2}
 \alpha = \arctan\left(\frac{a \cdot (q - L_{r})}{D \cdot (q + 1) \cdot (L_{r} + 1)}\right)
\end{equation}

\noindent where we have defined $\alpha$ so that $\alpha > 0$ signifies that the star dominates the observed reflex motion, i.e., $L_{r} < q$, and $\alpha < 0$ signifies that the planet dominates the observed reflex motion, i.e., $L_{r} > q$. Note that when the barycenter and photocenter are at the same point, i.e., $L_{r} = q$, and thus $\alpha = 0$, no reflex motion is observable.

We now estimate the value of $L_{r}$ based upon the values of readily measurable system parameters. Light emitted from the planet consists of both thermally emitted light, as well as incident stellar light reflected off the planet. Thus,

\begin{equation}
 L_{r} = \frac{L_{E} + L_{A}}{L_{\star}} = \frac{L_{E}}{L_{\star}} + \frac{L_{A}}{L_{\star}}
\end{equation}

\noindent where $L_{E}$ is the luminosity of the planet from thermal emission, L$_{\star}$ is the luminosity of the star, and $L_{A}$ is the luminosity of light reflected off the planet. To estimate the thermal component, we assume that both the star and planet radiate as blackbodies, and thus

\begin{equation}
 \label{thermaleq}
 \frac{L_{E}}{L_{\star}} = \frac{R_{p}^{2}}{R_{\star}^{2}} \cdot \frac{exp(\frac{hc}{\lambda k T_{\star}}) - 1}{exp(\frac{hc}{\lambda k T_{p}}) - 1}
\end{equation}

\noindent where $R_{p}$ is the radius of the planet, $\lambda$ is a given wavelength, $h$ is Planck's constant, $c$ is the speed of light, $k$ is Boltzmann's constant, $T_{\star}$ is the effective temperature of the star, and $T_{p}$ is the effective temperature of the planet. To derive $T_{p}$ we first assume that the planet is in radiative equilibrium, and has perfect heat re-distribution, i.e. a uniform planetary temperature, and thus

\begin{equation}
\label{tpeq}
 T_{p} = T_{\star} \cdot \left(\frac{(1-A_{B}) \cdot R_{\star}^{2}}{4 a^{2}}\right)^{\frac{1}{4}}
\end{equation}

\noindent where $T_{\star}$ is the temperature of the star, $A_{B}$ is the planetary Bond albedo, and $R_{\star}$ is the radius of the star. 

To estimate the contribution due to reflected light, we first note that the flux received at the planet's surface is $L_{\star}$ divided by the surface area of a sphere at a distance $a$, i.e., $4\pi a^{2}$. The planet intercepts and reflects this light on only one of its hemispheres, which has effective cross sectional area of $\pi R_{p}^{2}$, with an efficiency equal to the albedo. Combining these terms and re-arranging to obtain the luminosity ratio due to reflected light yields

\begin{equation}
\label{refleq}
 \frac{L_{A}}{L_{\star}} = \frac{A_{\lambda} R_{p}^{2}}{4a^{2}}
\end{equation}

\noindent where $A_{\lambda}$ is the planet's albedo at a given wavelength.

Combining the above equations, and assuming values of $A_{B}$ and $A_{\lambda}$, we can estimate $\alpha$ at a given $\lambda$, using only $M_{\star}$, $R_{\star}$, $T_{\star}$, $M_{p}$, $R_{p}$, $P$, and $D$. We note that this assumes that the planet is in radiative equilibrium, but does not account for any additional internal heat sources from the planet, such as gravitational contraction or radioactive decay. While internal heat sources are likely to be negligible for close-in planets, it could significantly contribute to the total luminosity of further out gaseous planets, thus making them even more easily detectable. Our approximation for $\alpha$ also assumes that the planet's luminosity is constant over its orbit as observed from Earth. However, some planets have significant flux differences between their day and night sides due to low day-to-night re-radiation efficiency and/or significant planetary albedos. In these cases, if the inclination of the system is $\neq$ 0$\degr$, then the planet's luminosity will vary with orbital phase as seen by the observer, and the projected astrometric orbit of the photocenter at wavelengths where the planet's luminosity dominates will deviate from an ellipse, with increasing deviation as the inclination approaches 90$\degr$. (This effect is further discussed and illustrated in Section~\ref{numsec}.) As well, we assumed a circular orbit, and thus eccentric planets with varying levels of stellar insolation and temperature would have unique orbital signatures resulting from time-variant planetary flux. Finally, we assumed in this analytical derivation that the star and planet are effectively point sources, but of course in reality they have a physical size. If the star and/or planet have non-symmetric surface features, such as star spots or planetary hot spots, then the star and planet could each influence the location of the photocenter as these features rotated across their surface. The effect on the photocenter would only be a fraction of their physical radii, and would only cause significant deviations to the observed astrometric orbit if the radii of either object was a significant fraction of the object's distance from the system's barycenter. While this would likely be negligible for the planet, it could be significant for the star, e.g., the case of microarcsecond, wavelength-dependent, astrometric perturbations resulting from star spots presented in Paper II.

\begin{figure*}
\centering
\begin{tabular}{ccc}
\epsfig{width=0.3\linewidth,file=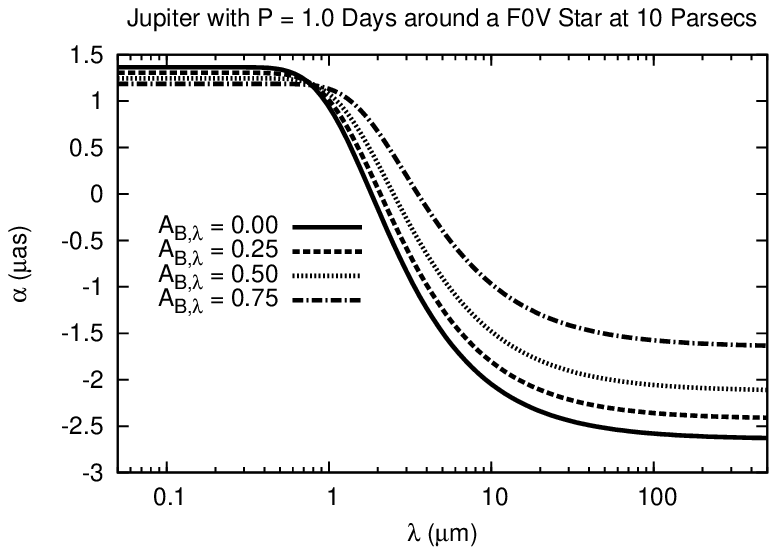} &
\epsfig{width=0.3\linewidth,file=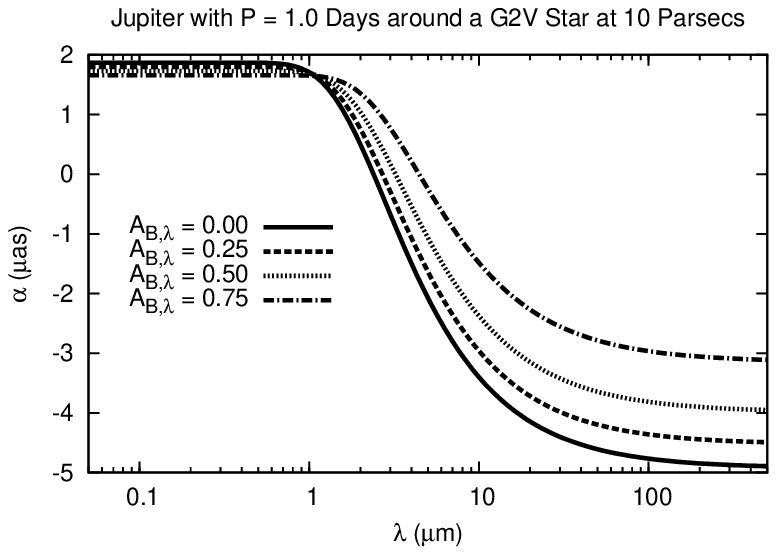} &
\epsfig{width=0.3\linewidth,file=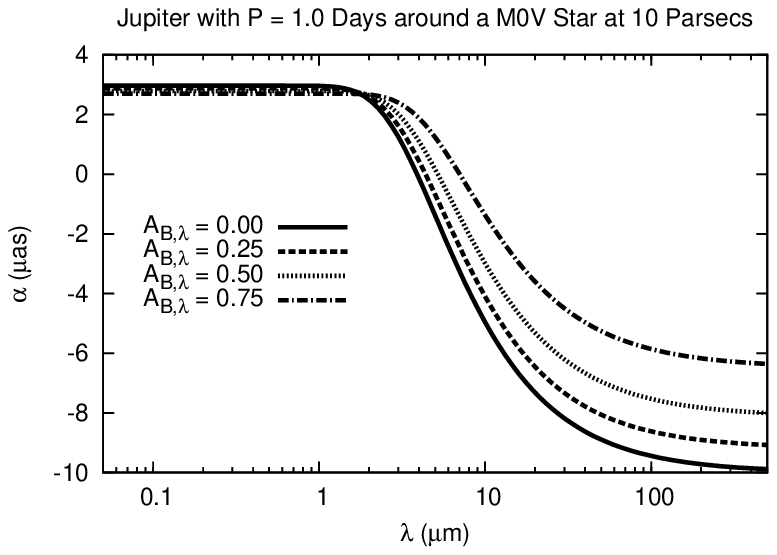} \\
\epsfig{width=0.3\linewidth,file=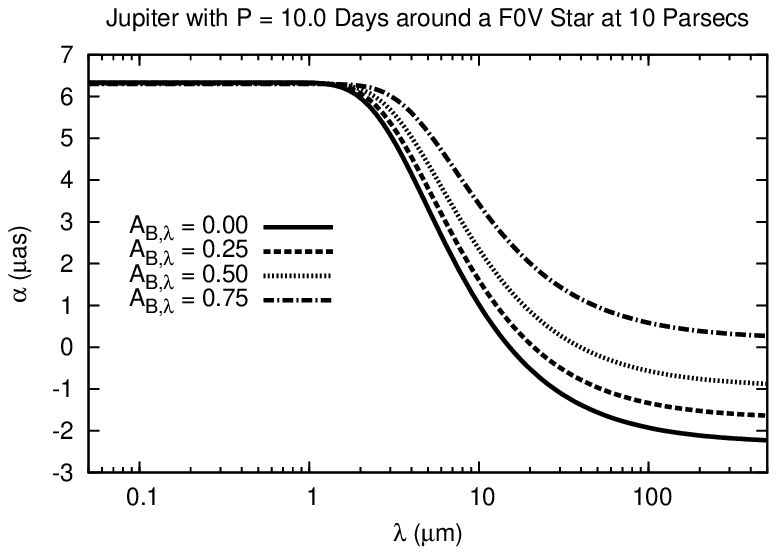} &
\epsfig{width=0.3\linewidth,file=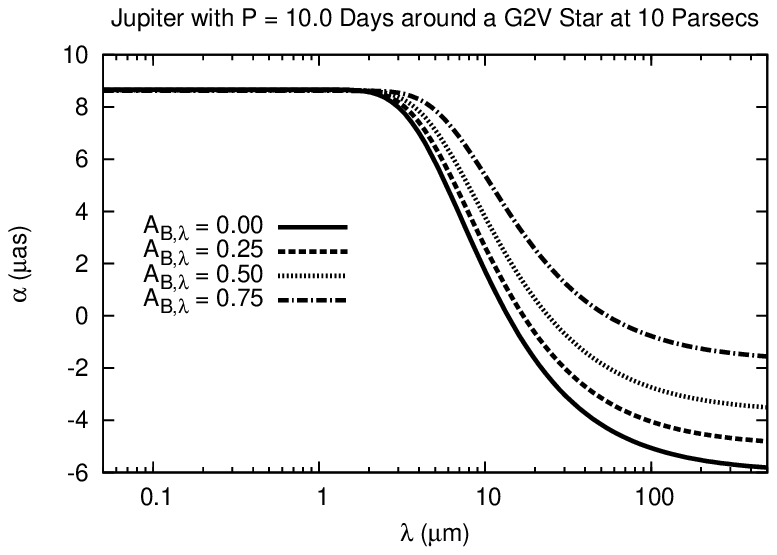} &
\epsfig{width=0.3\linewidth,file=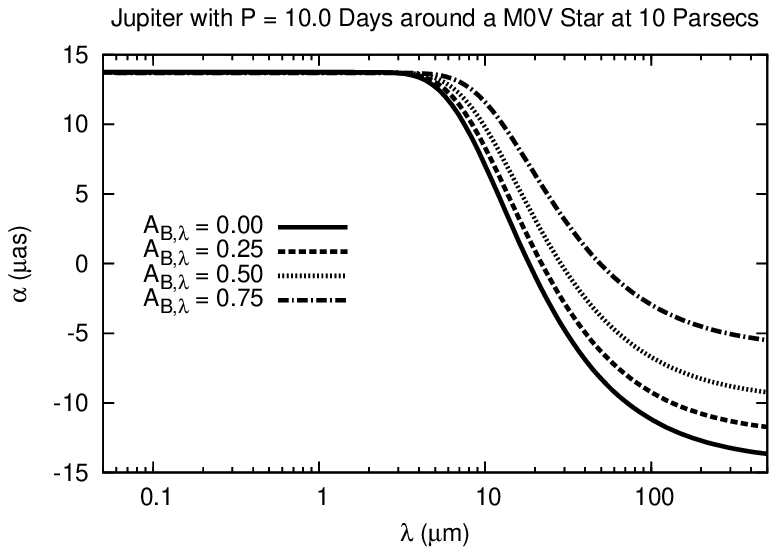} \\
\epsfig{width=0.3\linewidth,file=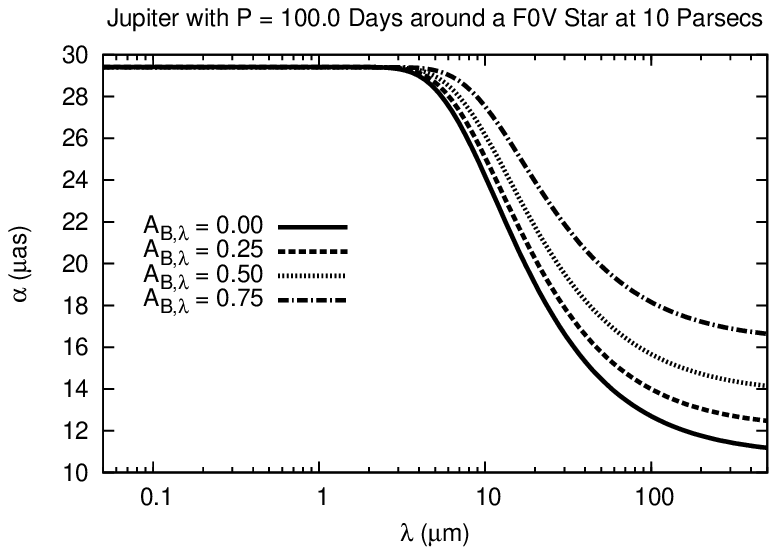} &
\epsfig{width=0.3\linewidth,file=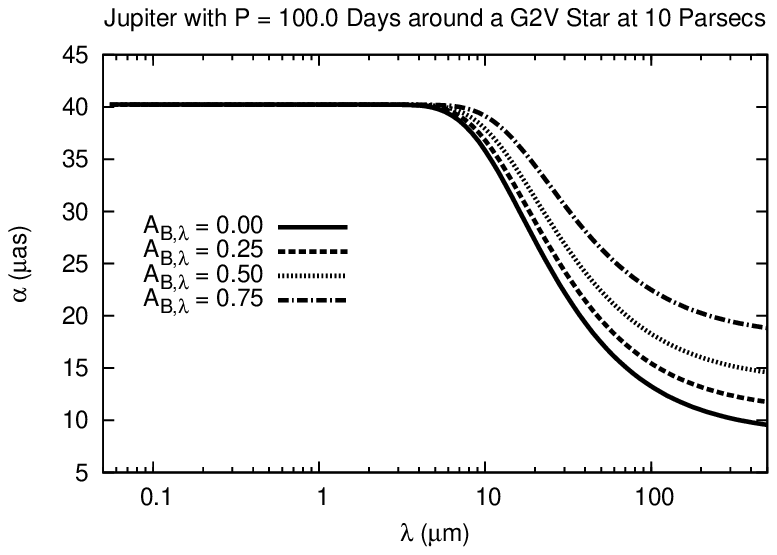} &
\epsfig{width=0.3\linewidth,file=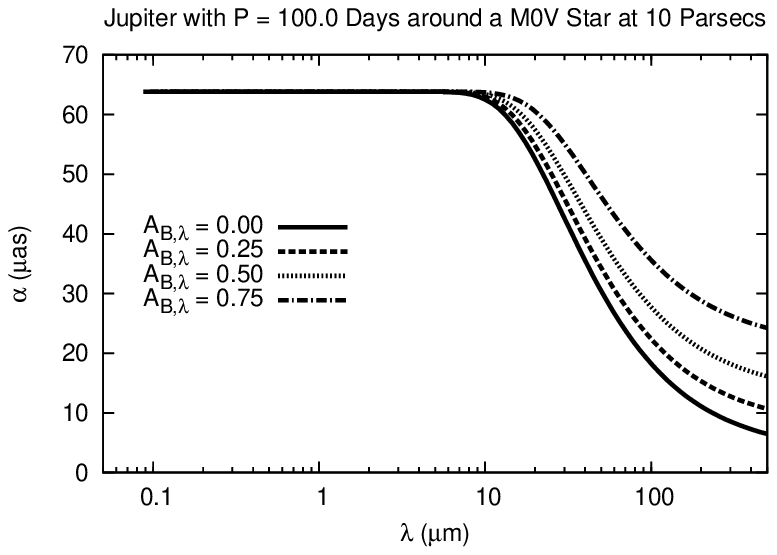} \\
\epsfig{width=0.3\linewidth,file=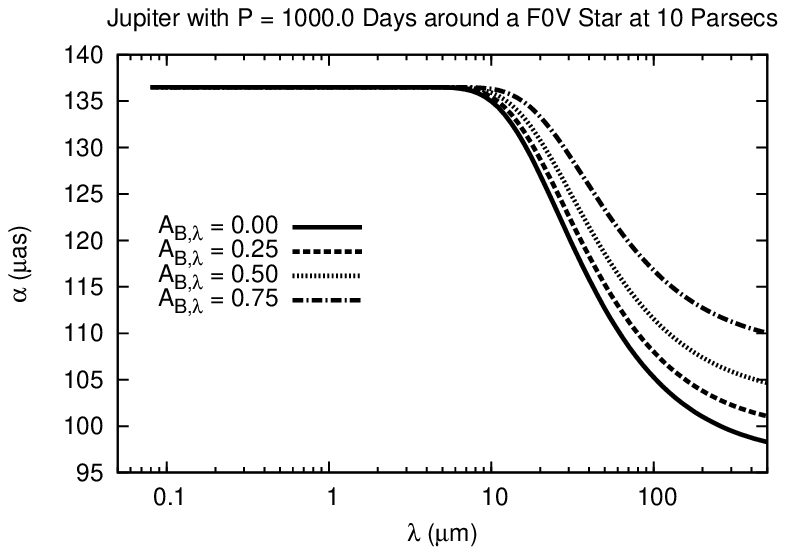} &
\epsfig{width=0.3\linewidth,file=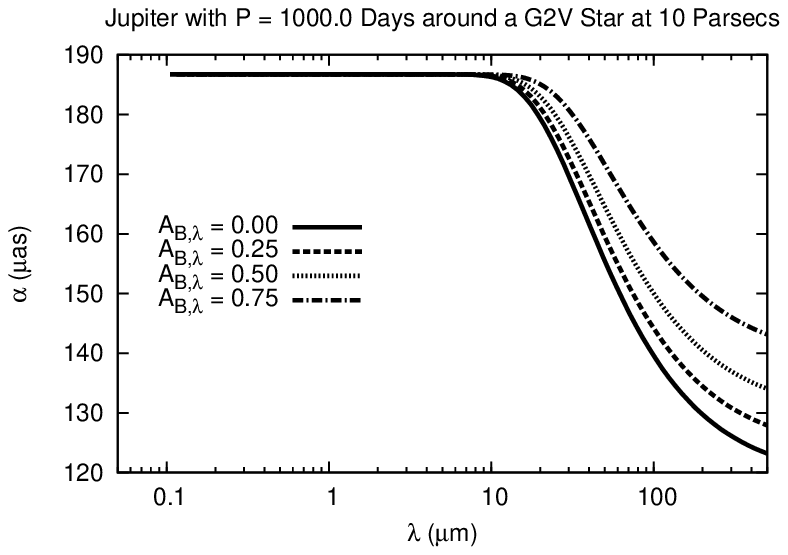} &
\epsfig{width=0.3\linewidth,file=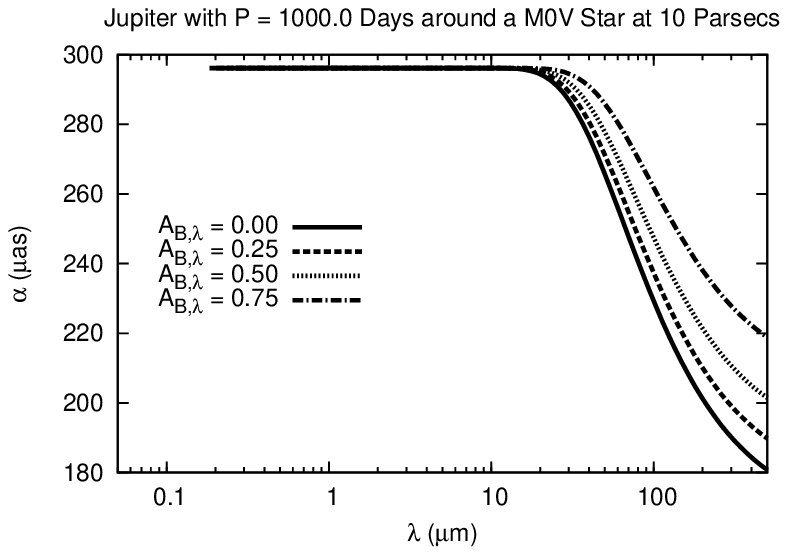} \\
\end{tabular}
\caption{Plots of the reflex motion amplitude, $\alpha$, versus the wavelength of observations, $\lambda$, for a Jupiter-like planet, ($M_{p}$ = 1.0 $M_{J}$, $R_{p}$ = 1.0 $R_{J}$) around F0V, G2V, and M0V stars at 10 parsecs, (left, middle, and right columns respectively), at periods of 1, 10, 100, and 1000 days, (top to bottom rows, respectively). The solid, dashed, dotted, and dash-dotted lines represent planetary albedos of 0.0, 0.25, 0.5, and 0.75 respectively.}
\label{jupsimfig}
\end{figure*}

\begin{figure*}
\centering
\begin{tabular}{ccc}
\epsfig{width=0.3\linewidth,file=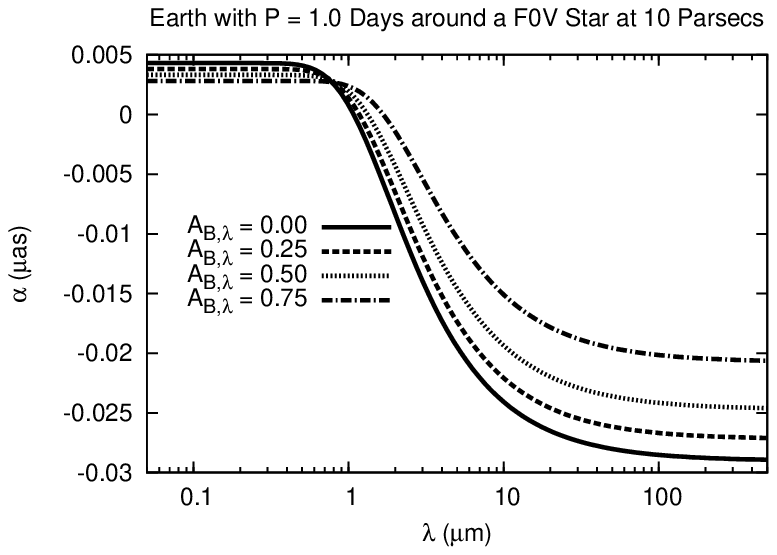} &
\epsfig{width=0.3\linewidth,file=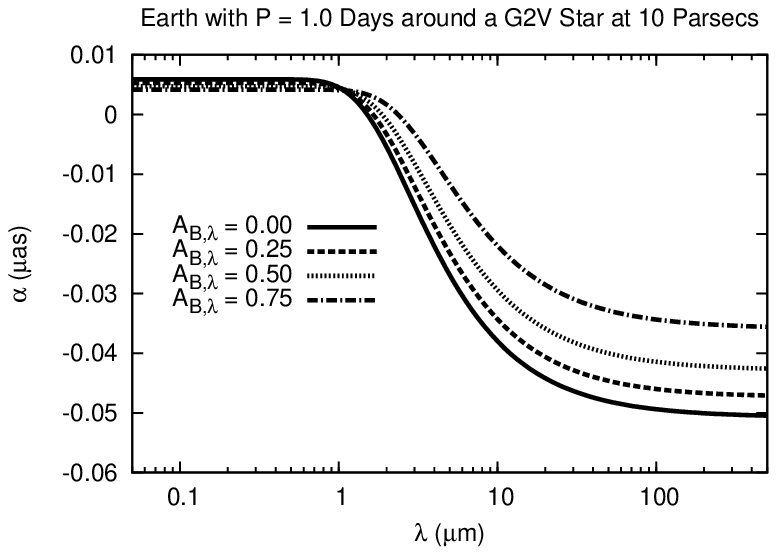} &
\epsfig{width=0.3\linewidth,file=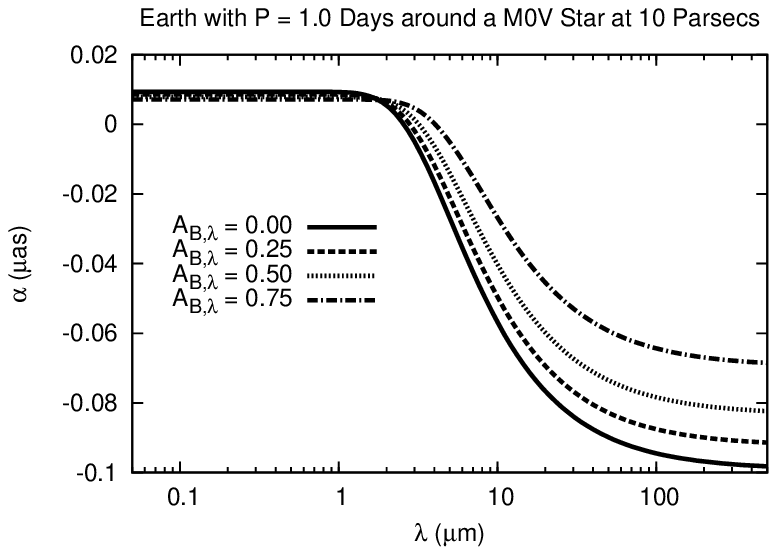} \\
\epsfig{width=0.3\linewidth,file=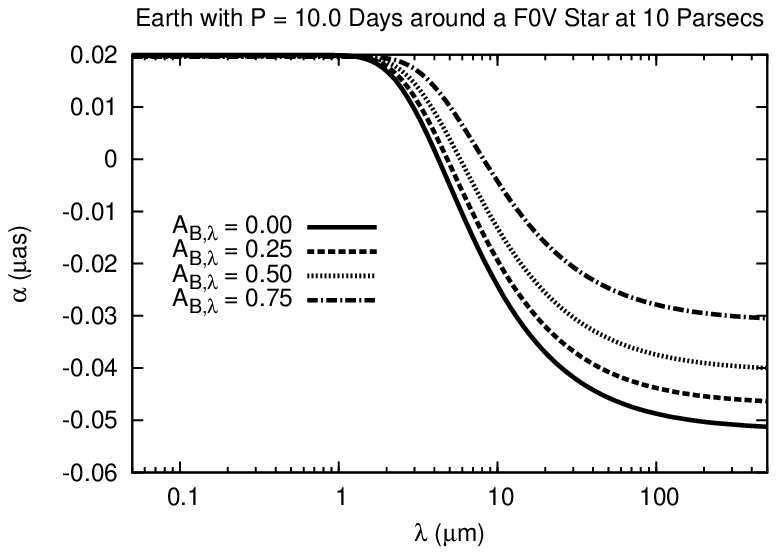} &
\epsfig{width=0.3\linewidth,file=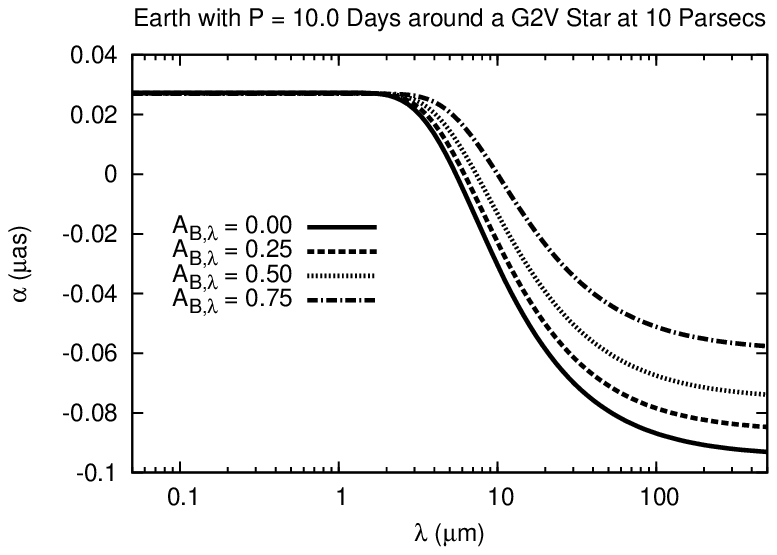} &
\epsfig{width=0.3\linewidth,file=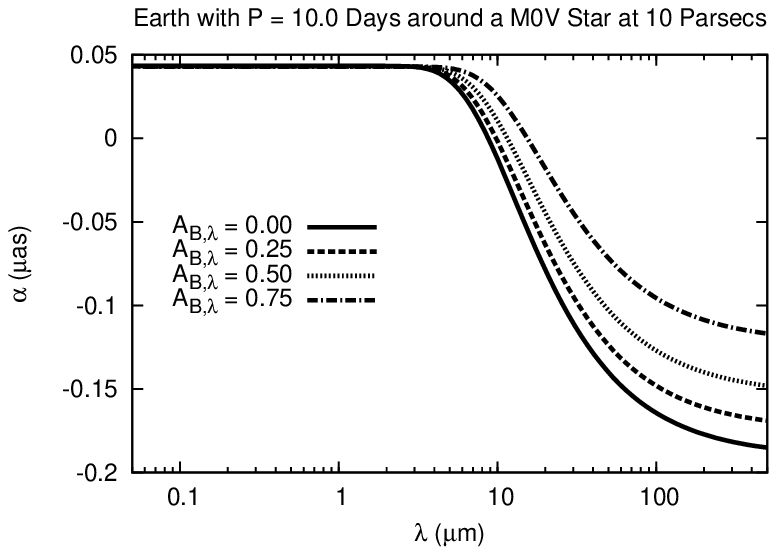} \\
\epsfig{width=0.3\linewidth,file=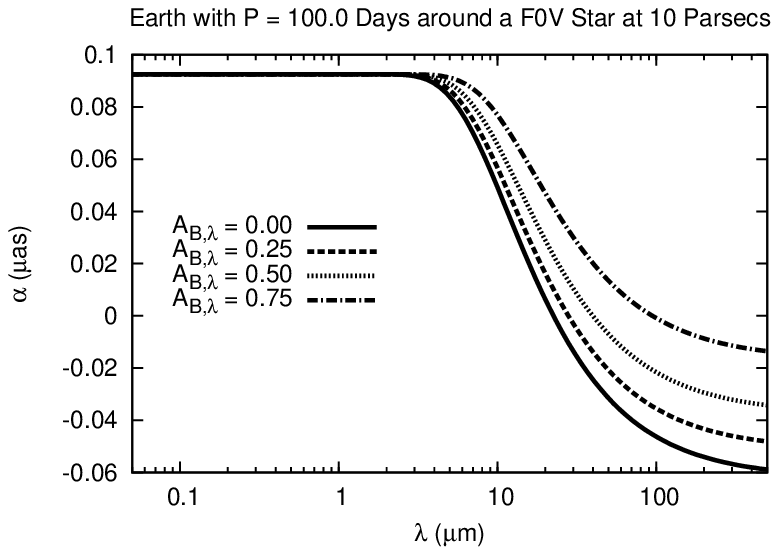} &
\epsfig{width=0.3\linewidth,file=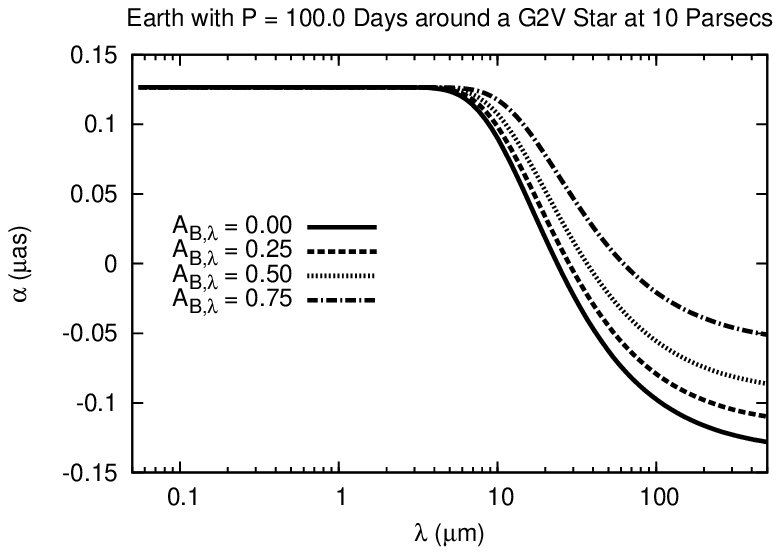} &
\epsfig{width=0.3\linewidth,file=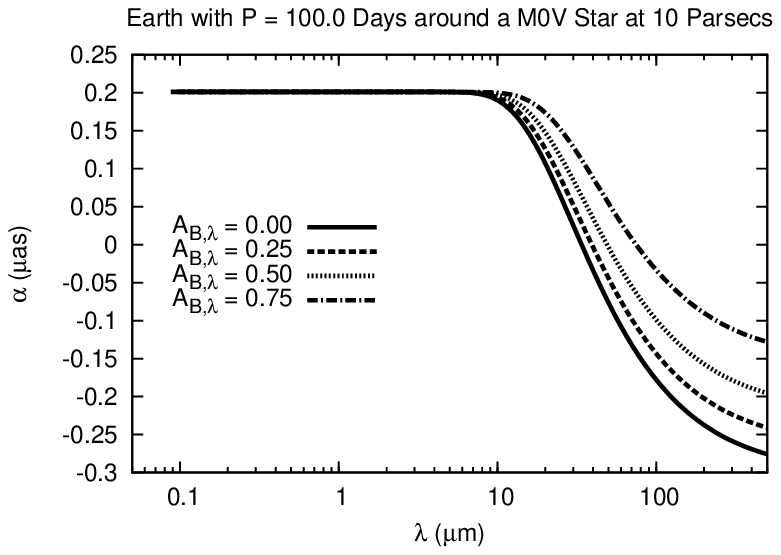} \\
\epsfig{width=0.3\linewidth,file=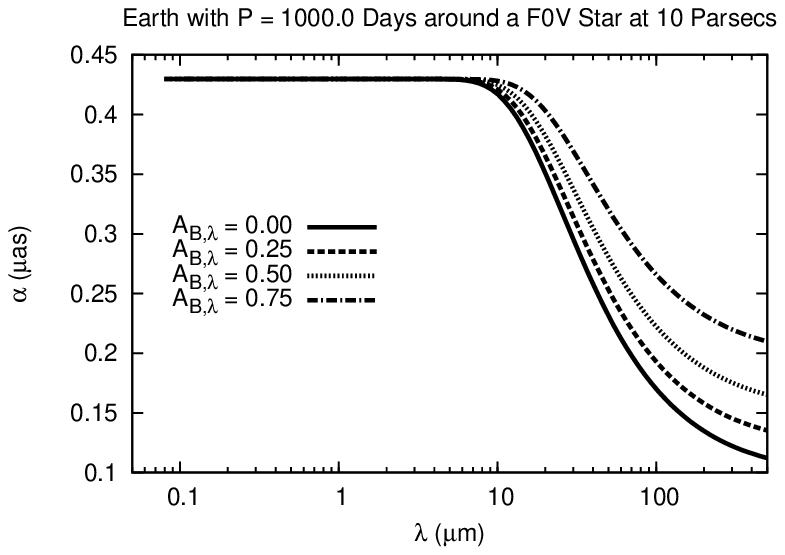} &
\epsfig{width=0.3\linewidth,file=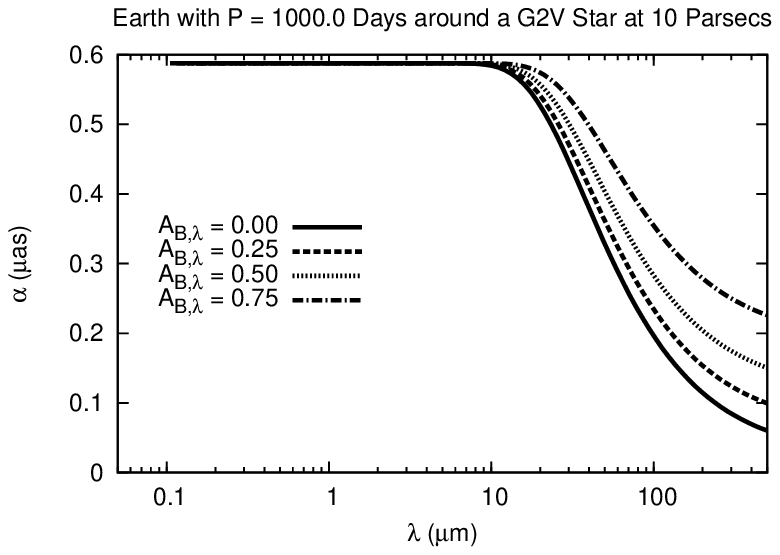} &
\epsfig{width=0.3\linewidth,file=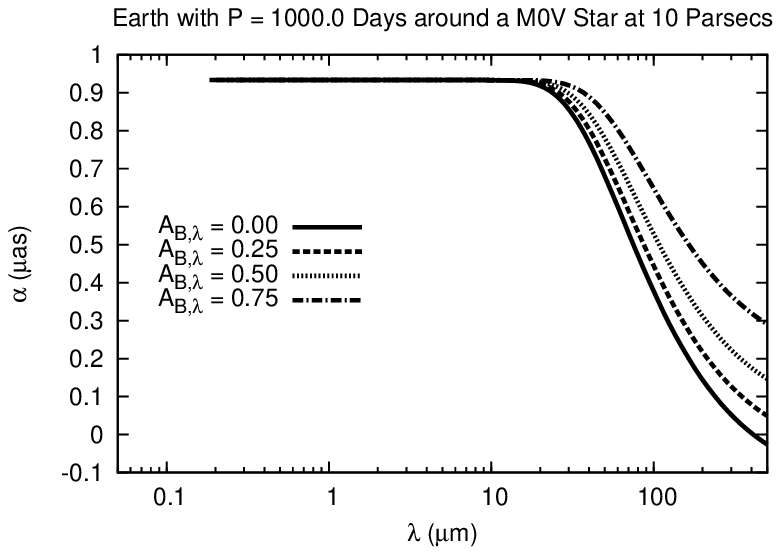} \\
\end{tabular}
\caption{Plots of the reflex motion amplitude, $\alpha$, versus the wavelength of observations, $\lambda$, for an Earth-like planet, ($M_{p}$ = 1.0 $M_{\earth}$, $R_{p}$ = 1.0 $R_{\earth}$) around F0V, G2V, and M0V stars at 10 parsecs, (left, middle, and right columns respectively), at periods of 1, 10, 100, and 1000 days, (top to bottom rows, respectively). The solid, dashed, dotted, and dash-dotted lines represent planetary albedos of 0.0, 0.25, 0.5, and 0.75 respectively.}
\label{earthsimfig}
\end{figure*}

In Figure~\ref{jupsimfig} we present plots of $\alpha$ versus $\lambda$ for a Jupiter-like planet, ($M_{p}$ = 1.0 $M_{J}$, $R_{p}$ = 1.0 $R_{J}$), around F0V, G2V, and M0V stars at 10 parsecs, with periods of 1, 10, 100, and 1000 days. We also show various planetary albedos, assuming $A_{B}$ = $A_{\lambda}$, of 0.0, 0.25, 0.5, and 0.75. In Figure~\ref{earthsimfig} we do the same for an Earth-like planet, ($M_{p}$ = 1.0 $M_{\earth}$, $R_{p}$ = 1.0 $R_{\earth}$). In general, systems that have large, high-mass stars and large, low-mass planets present the best opportunity to observe negative values of $\alpha$, and thus be able to directly determine their masses. (This is a unique parameter space not covered by other exoplanet characterization techniques such as radial-velocity or the transit method.) Short-period, and thus hot, planets around more massive stars transition to negative values of $\alpha$ at shorter wavelengths, but have lower overall amplitudes compared to long-period, and thus cool, planets around low-mass stars. Reflected light is a fairly minor contribution, only having some significant relevance for planets with very short orbital periods, i.e., $\sim$1 day. For both hot Jupiters and hot Earths, negative values of $\alpha$ can be observed with wavelengths as short as $\sim$2 $\mu$m, i.e., the K band. Considering $\lambda$ $<$ 100 $\mu$m, $\alpha$ $<$ 0 could only be observed for $P$ $\lesssim$ 100 days for a Jupiter-like planet, and for $P$ $\lesssim$ 500 days for an Earth-like planet. Earth itself, (P = 365 days around a G2V star), would have a value of $\alpha$ $\approx$ 0.3 $\mu$as for $\lambda$ $<$ 10 $\mu$m, and $\alpha$ $\approx$ -0.05 $\mu$as at 100 $\mu$m, and thus, theoretically, the absolute mass of an Earth-analogue and its host star could be determined via this technique.

Utilizing exoplanets.org, we have collected the values for all the previously mentioned system parameters for all currently known exoplanets. Selecting those that have well-determined values of all the needed parameters, in Table~\ref{tab1} we list the top five exoplanets with the largest negative values of $\alpha$ for each of the K (2.19 $\mu$m), L (3.45 $\mu$m), M (4.75 $\mu$m), and N (10.0 $\mu$m) infrared bandpasses, with a total of 11 unique exoplanets. We choose these wavelengths as they are the major ground-based infrared observing windows, and no systems examined had negative $\alpha$ values at wavelengths shorter than $\sim$2 $\mu$m. All of the candidate systems ended up being transiting planets both because they have well-determined values for the planetary radii, and transit surveys are most sensitive to close-in planets. As can be seen, the top candidates for detecting $\alpha$ $<$ 0, and thus measuring the absolute mass of the planet, are WASP-12 b in the K band with $\alpha$ = -0.05 $\mu$as, HD 209458 b in the L and M bands with $\alpha$ = -0.23 and -0.66 $\mu$as respectively, and HD 189733 b in the N band with $\alpha$ = -3.04 $\mu$as. It is interesting that three low-mass Neptune and sub-Neptune mass planets, 55 Cnc e, Gliese 436 b, and GJ 1214 b, also make the list, illustrating that this technique can `favor' the characterization of low-mass planets.

\tabletypesize{\scriptsize}
\setlength{\tabcolsep}{0.105cm}
\begin{deluxetable}{ccccccccc}
  \tablewidth{0pt}
  \tablecaption{Currently Known Exoplanets with the Most Negative $\alpha$ Values}
  \tablecolumns{9}
  \tablehead{Name & $D$ & $M_{\star}$ & $R_{\star}$ & $T_{\star}$ & $M_{p}$ & $R_{p}$ & $P$ & $\alpha$ \\ & (pc) & (M$_{\sun}$) & (R$_{\sun}$) & (K) & (M$_{\rm J}$) & (R$_{\rm J}$) & (Days) & ($\mu$as)}
  \startdata
  \cutinhead{K Band (2.19 $\mu$m)}
  WASP-12 b & 427 & 1.28 & 1.63 & 6300 & 1.35 & 1.79 & 1.091 & -0.05 \\ 
WASP-19 b & 250 & 0.93 & 0.99 & 5500 & 1.11 & 1.39 & 0.789 & -0.05 \\ 
WASP-33 b & 115 & 1.50 & 1.44 & 7430 & 2.05 & 1.50 & 1.220 & -0.04 \\ 
55 Cnc e & 12 & 0.96 & 0.96 & 5234 & 0.03 & 0.19 & 0.737 & -0.01 \\ 
CoRoT-1 b & 480 & 0.95 & 1.11 & 5950 & 1.03 & 1.49 & 1.509 & -0.01 \\

  \cutinhead{L Band (3.45 $\mu$m)}
  HD 209458 b & 49 & 1.13 & 1.16 & 6065 & 0.69 & 1.36 & 3.525 & -0.23 \\ 
WASP-33 b & 115 & 1.50 & 1.44 & 7430 & 2.05 & 1.50 & 1.220 & -0.20 \\ 
WASP-19 b & 250 & 0.93 & 0.99 & 5500 & 1.11 & 1.39 & 0.789 & -0.15 \\ 
WASP-17 b & 300 & 1.19 & 1.20 & 6550 & 0.49 & 1.51 & 3.735 & -0.11 \\ 
WASP-12 b & 427 & 1.28 & 1.63 & 6300 & 1.35 & 1.79 & 1.091 & -0.10 \\ 

  \cutinhead{M Band (4.75 $\mu$m)}
  HD 209458 b & 49 & 1.13 & 1.16 & 6065 & 0.69 & 1.36 & 3.525 & -0.66 \\ 
HD 189733 b & 19 & 0.81 & 0.76 & 5040 & 1.14 & 1.14 & 2.219 & -0.47 \\ 
WASP-33 b & 115 & 1.50 & 1.44 & 7430 & 2.05 & 1.50 & 1.220 & -0.29 \\ 
WASP-19 b & 250 & 0.93 & 0.99 & 5500 & 1.11 & 1.39 & 0.789 & -0.21 \\ 
WASP-17 b & 300 & 1.19 & 1.20 & 6550 & 0.49 & 1.51 & 3.735 & -0.19 \\ 

  \cutinhead{N Band (10.0 $\mu$m)}
  HD 189733 b & 19 & 0.81 & 0.76 & 5040 & 1.14 & 1.14 & 2.219 & -3.04 \\ 
HD 209458 b & 49 & 1.13 & 1.16 & 6065 & 0.69 & 1.36 & 3.525 & -1.53 \\ 
Gliese 436 b & 10 & 0.45 & 0.46 & 3684 & 0.07 & 0.38 & 2.644 & -0.95 \\ 
WASP-34 b & 120 & 1.01 & 0.93 & 5700 & 0.58 & 1.22 & 4.318 & -0.64 \\ 
GJ 1214 b & 12 & 0.16 & 0.21 & 3026 & 0.02 & 0.24 & 1.580 & -0.59 

  \enddata
  \label{tab1}
\end{deluxetable}

\section{Numerical Modeling via \sc{reflux}}
\label{numsec}

In order to provide a check on our analytical formulae, better illustrate the multi-wavelength astrometric orbits of exoplanet systems, and probe some more subtle effects, we use the \textsc{reflux}\footnotemark[3] code \citep{Coughlin2010a}, which computes the flux-weighted astrometric reflex motion of binary systems at multiple wavelengths, to model a couple known exoplanet systems. We discussed the code in detail in Papers I and II, but in short, it utilizes the Eclipsing Light Curve (ELC) code, which was written to compute light curves of eclipsing binary systems \citep{Orosz2000}. ELC includes the dominant physical effects that shape a binary system's light curve, such as non-spherical geometry due to rotation and tidal forces, gravity darkening, limb darkening, mutual heating, reflection effects, and the inclusion of hot or cool spots on the stellar surface. The ELC code represents the surfaces of two stars, or a star-planet system, as a grid of individual luminosity points, and calculates the resulting light curve given the provided systemic parameters. \textsc{reflux} takes the grid of luminosity points at each phase and calculates the flux-weighted astrometric photocenter location at each phase, taking into account the system's distance from Earth. Although ELC is capable of using model atmospheres, for this paper we set the code to calculate luminosities assuming both the star and planet radiate as blackbodies.

\footnotetext[3]{\textsc{reflux} can be run via a web interface from \url{http://astronomy.nmsu.edu/jlcough/reflux.html}. Additional details as to how to set-up a model are presented there.}

We choose to model Wasp-12, HD 209458, and HD 189733, as they are all well-studied systems, and have the most negative $\alpha$ values for the K, L, M, and N bandpasses presented in Table~\ref{tab1}. For each system we set the values for $M_{\star}$, $R_{\star}$, $T_{\star}$, $M_{p}$, $R_{p}$, $P$, $D$, and rotation period of the star to those in the Exoplanets.org database, and set the rotation period of the planet to the orbital period of the system, i.e., assume the planet is tidally locked, and assume a circular orbit. We assume that the spin axes of both the star and planet are perfectly aligned with the orbital axis. We employ the use of spots in the ELC code to simulate a day/night side temperature difference, by assuming a uniform day-side temperature for the planetary hemisphere facing the star, and a uniform night-side temperature for the planetary hemisphere facing away from the star. We employ the values for the day and night side temperatures derived by \citet{Cowan2011}, which were 2939 K for the day-side of Wasp-12 b, 1486 and 1476 K for the day and night sides respectively of HD 209458 b, and 1605 and 1107 K for the day and night sides respectively of HD 189733 b. We adopted a temperature of 1470 K for the night side of Wasp-12 b, i.e., half that of the day side, assuming very little planetary heat redistribution. For all the systems, we set the star's gravity darkening coefficients to those determined by \citet{Claret2000}, though do not enable gravity darkening for the planet. For both the planet and star, we assume zero albedo, since we are dealing principally with infrared wavelengths where the effect is negligible, and we have already shown that even in the optical reflected light is a minor contribution to the astrometric motions under investigation. Furthermore, the chosen planets are expected to have very low albedos ($A_{B} <$ 0.3) from model atmospheres \citep{Marley1999,Seager2000,Sudarsky2000}, and have even had their albedos constrained to very low values from observations, e.g., \citet{LopezMorales2010} for Wasp-12b, \citet{Rowe2008} for HD 209458b, and \citet{Wiktorowicz2009} for HD 189733b. We also do not assume any limb-darkening since we are dealing principally with infrared wavelengths.

\begin{figure*}
\centering
\begin{tabular}{cc}
\epsfig{width=0.45\linewidth,file=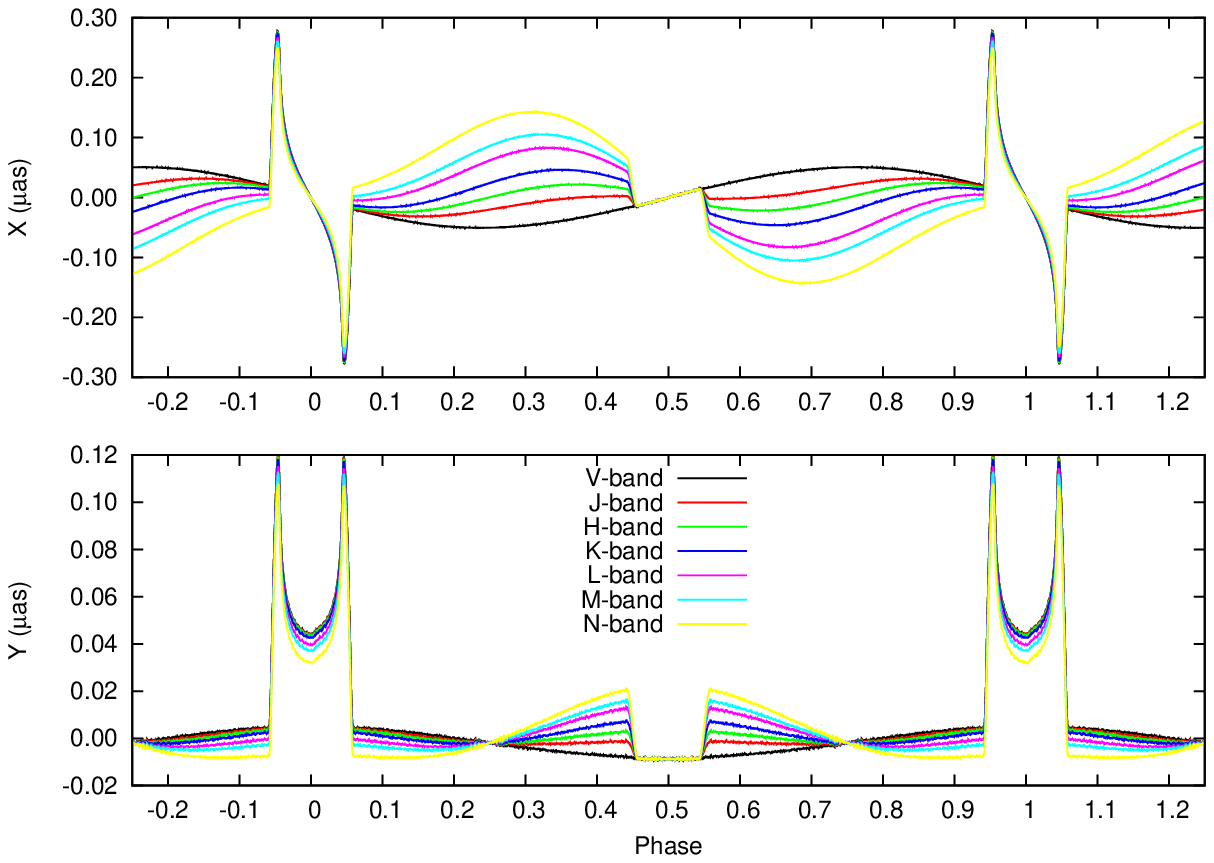} &
\epsfig{width=0.45\linewidth,file=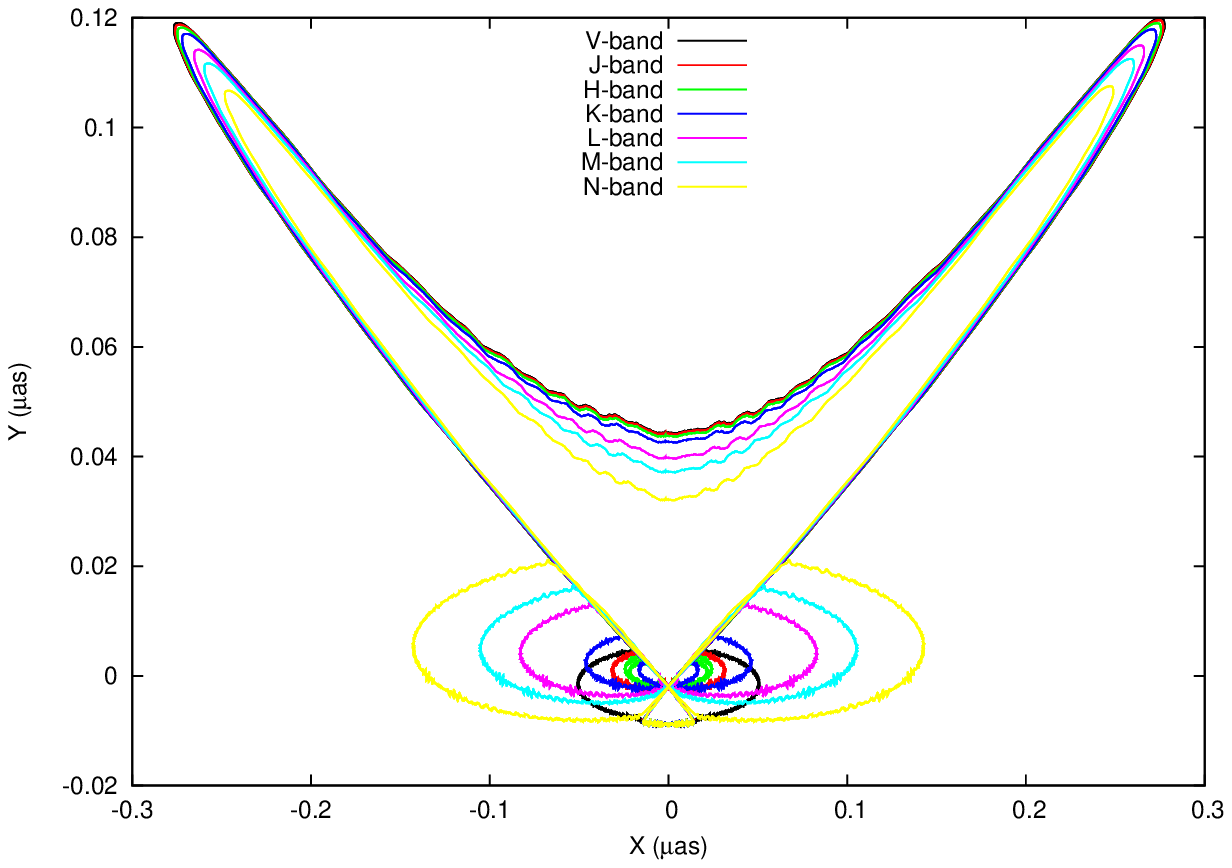} \\
\end{tabular}
\caption{Plots of the multi-wavelength astrometric orbit for the Wasp-12 system. Left: The X and Y components of motion versus phase. Right: The sky-projected, X-Y, orbit. The point (X,Y) = (0,0) corresponds to the system's barycenter, and the projected orbital rotation axis is parallel to the Y-axis. Phase 0.0 corresponds to the primary transit, when the planet passes in front of the star and is closest to the observer, and phase 0.5 corresponds to the secondary eclipse, when the planet passes behind the star and is farthest away from the observer.}
\label{refluxmodel1}
\end{figure*}

\begin{figure*}
\centering
\begin{tabular}{cc}
\epsfig{width=0.45\linewidth,file=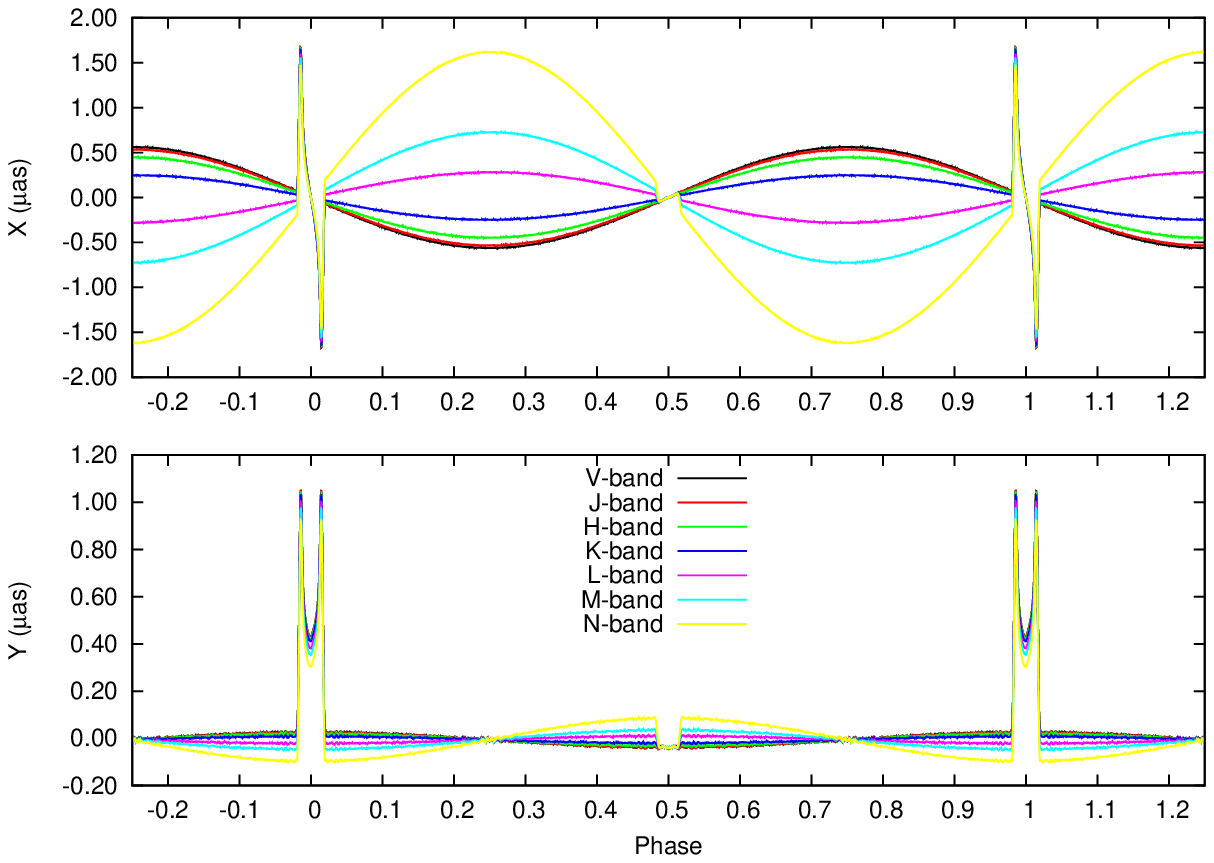} &
\epsfig{width=0.45\linewidth,file=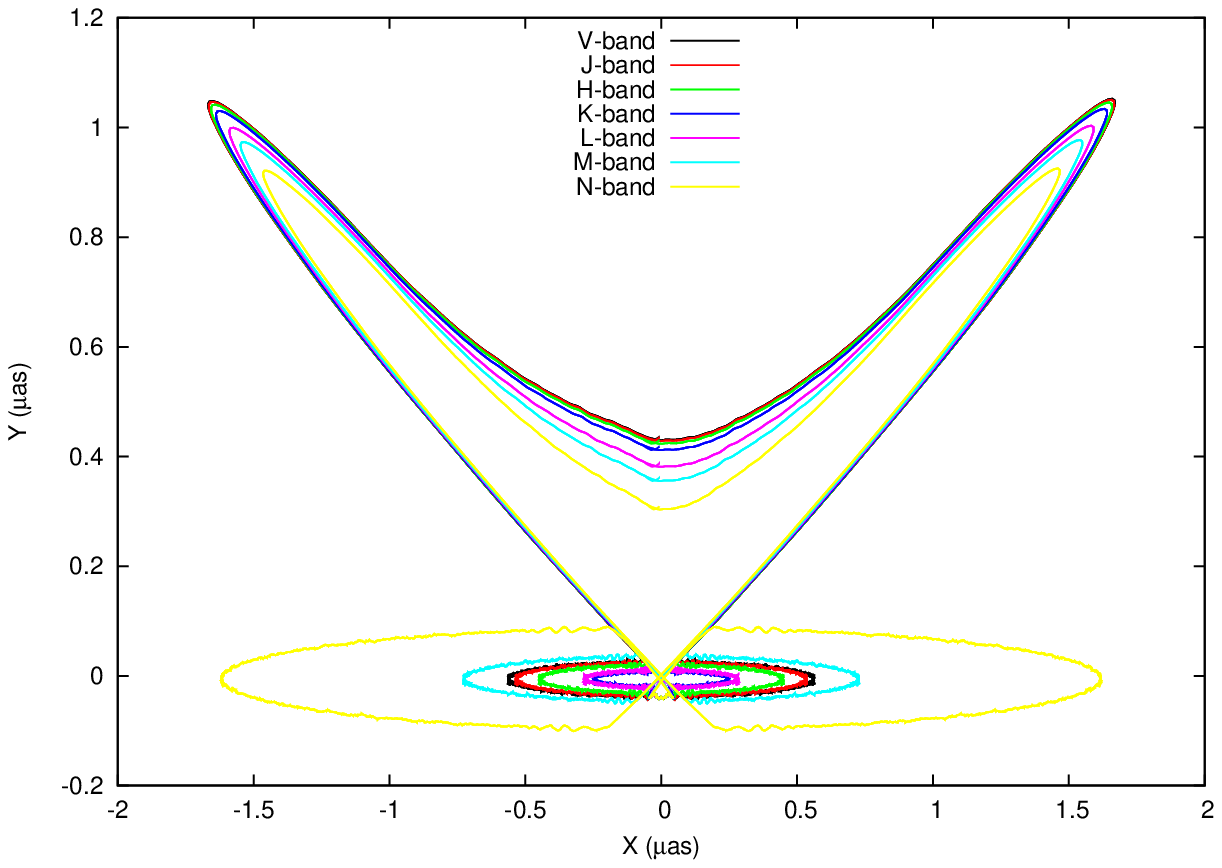} \\
\end{tabular}
\caption{Plots of the multi-wavelength astrometric orbit for the HD 209458 system. Left: The X and Y components of motion versus phase. Right: The sky-projected, X-Y, orbit. The point (X,Y) = (0,0) corresponds to the system's barycenter, and the projected orbital rotation axis is parallel to the Y-axis. Phase 0.0 corresponds to the primary transit, when the planet passes in front of the star and is closest to the observer, and phase 0.5 corresponds to the secondary eclipse, when the planet passes behind the star and is farthest away from the observer.}
\label{refluxmodel2}
\end{figure*}

\begin{figure*}
\centering
\begin{tabular}{cc}
\epsfig{width=0.45\linewidth,file=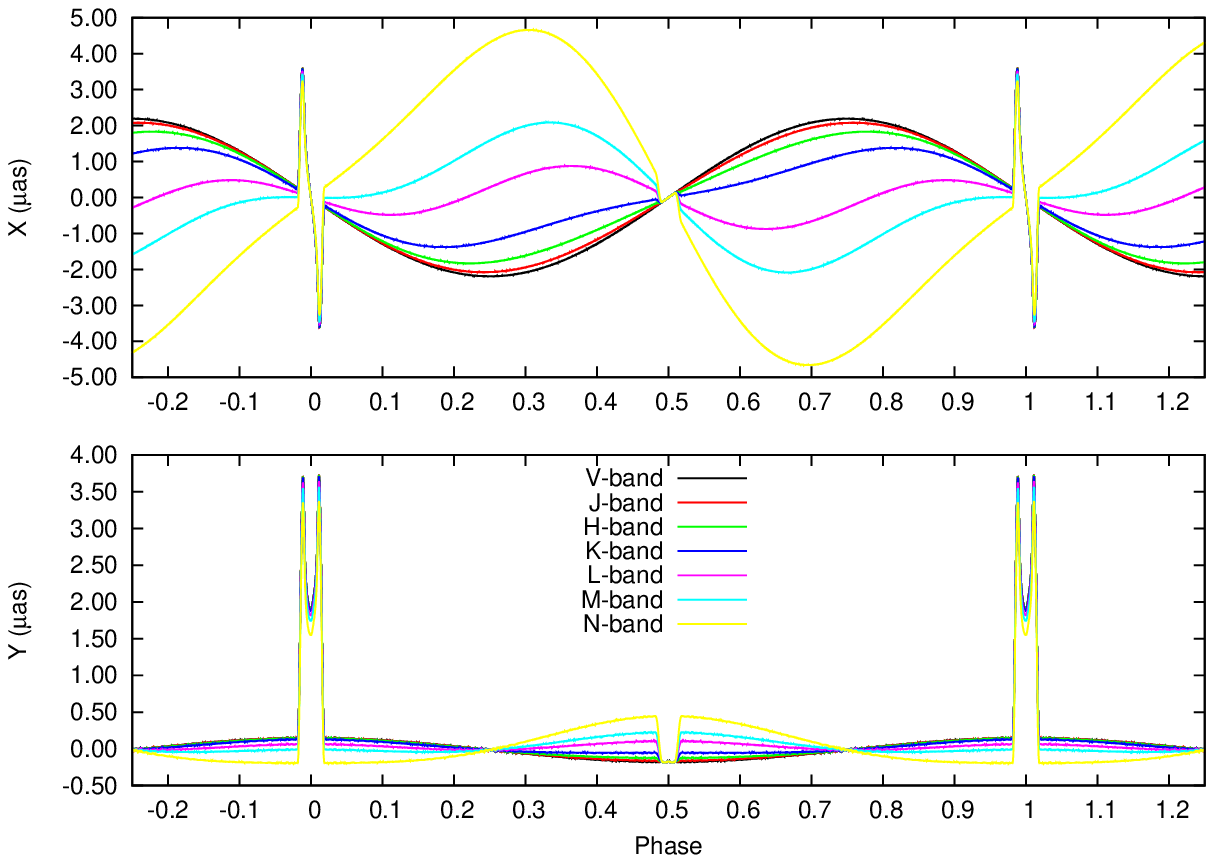} &
\epsfig{width=0.45\linewidth,file=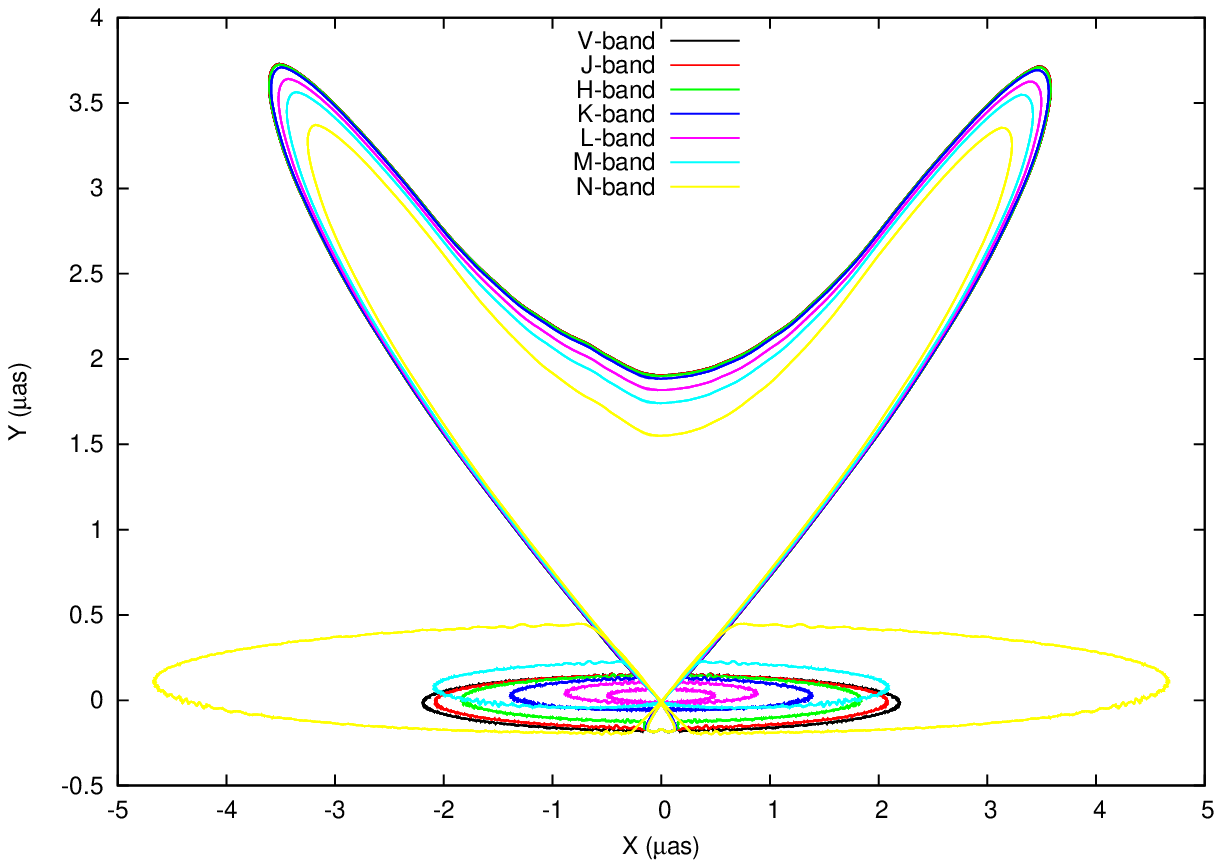} \\
\end{tabular}
\caption{Plots of the multi-wavelength astrometric orbit for the HD 189733 system. Left: The X and Y components of motion versus phase. Right: The sky-projected, X-Y, orbit. The point (X,Y) = (0,0) corresponds to the system's barycenter, and the projected orbital rotation axis is parallel to the Y-axis. Phase 0.0 corresponds to the primary transit, when the planet passes in front of the star and is closest to the observer, and phase 0.5 corresponds to the secondary eclipse, when the planet passes behind the star and is farthest away from the observer.}
\label{refluxmodel3}
\end{figure*}

In Figures~\ref{refluxmodel1}, \ref{refluxmodel2}, and \ref{refluxmodel3} we present plots of the X and Y components of the photocenter versus phase, as well as the sky-projected X-Y orbit of the photocenter, in the V, J, H, K, L, M, and N passbands, for Wasp-12, HD 209458, and HD 189733 respectively. The point (X,Y) = (0,0) corresponds to the barycenter of the system, and the projected orbital rotation axis is parallel to the Y-axis. Phase 0.0 corresponds to the primary transit, when the planet passes in front of the star and is closest to the observer, and phase 0.5 corresponds to the secondary eclipse, when the planet passes behind the star and is farthest away from the observer.

Examining the modeling results, the values for $\alpha$ determined via the analytical formulae appear to match the numerical modeling results fairly well. For example, via Table~\ref{tab1}, Wasp-12b was predicted to have $\alpha$ values of -0.05 and -0.10 $\mu$as in the K and L bands respectively, compared to the maximum, out-of-transit, numerical model results of -0.05 and -0.08 $\mu$as. For HD 209458b, expected $\alpha$ values were -0.23, -0.66, and -1.53 $\mu$as for the L, M, and N bands, compared to -0.30, -0.74, and -1.63 $\mu$as from the numerical models. For HD 189733b, expected $\alpha$ values were -0.47 and -3.04 $\mu$as for the M and N bands, compared to -2.10 and -4.68 $\mu$as from the numerical models. The differences are principally due to the use of observationally determined day and night side temperatures in the numerical models, whereas the analytical formulae assumed perfect radiative equilibrium and a uniform planetary temperature.

Although a transition from positive to negative $\alpha$ appears to occur around the H, K, and L bands for Wasp-12 b, HD 209458 b, and HD 189733 respectively, a deviation from the visible light signature is clearly visible at shorter wavelengths, and thus it may be possible to disentangle the astrometric motion due to the planet even at shorter wavelengths where it does not dominate the reflex motion of the photocenter. For Wasp-12 b and HD 189733 b the out of transit/eclipse signature deviates from a sinusoid due to the extreme day/night temperature differences on these planets, as discussed in Section~\ref{eqsec}. The different inclinations of the systems are immediately apparent in the X-Y orbit plots, and when actually measured on sky, would directly yield the three-dimensional orbit of the system.

The presence of the primary transit and secondary eclipse is clearly visible in all three cases, with the primary transit dominating the maximum amplitude of the astrometric shift for the visible wavelengths, particularly in the Y-direction. As no limb-darkening was assumed in these models, the variation in the primary and secondary eclipse signatures with wavelength is due to the relative flux of the star and planet in those passbands. As noted by \citet{Gaudi2010}, measuring the astrometric shift of the primary transit directly yields the angular radius of the host star, and if the distance to the system is precisely known, one can directly derive the physical radius of the star. Additionally, if the density of the star is directly determined from the photometric light curve \citep{SeagerOrnelas2003}, then one can also directly derive the mass of the star. We also note, for the first time, that measuring the astrometric signature of the primary transit and, if observing at longer wavelengths, the secondary eclipse, specifically the duration of ingress and egress, similarly directly yields the angular radius of the planet. Since one may directly determine the surface gravity of the planet from the photometric light and radial-velocity curves alone \citep*{Southworth2007}, one may also directly determine the mass of the planet. Thus, for transiting planets, multi-wavelength astrometric measurements yield two independent methods of measuring the physical stellar and planetary masses.

\section{Discussion and Summary}
\label{discusssec}

We have shown that the multi-wavelength astrometric measurements of exoplanetary systems can be used to directly determine the masses of extrasolar planets and their host stars, in addition to the inclination and spatial orientation of their orbital axis. If the planet happens to transit the host star, then the angular radius of both the star and planet can be directly determined, and when combined with the trigonometric parallax of the system, the absolute radii of the planet and host star can directly determined via astrometry alone. We found that this technique is best suited, though is certainly not limited to, large, low-mass planets that orbit large, high-mass stars, and thus covers a unique parameter space not usually covered by other exoplanet characterization techniques. 

We have provided analytical formulae and numerical models to estimate the amplitude of the photocenter motion at various wavelengths. We found that, for some systems, the planet can dominate the motion of the system's photocenter at wavelengths as short as $\sim$2 $\mu$m, though the amplitude of the effect is only $\sim$0.05 $\mu$as. If one is able to obtain astrometric measurements at wavelengths up to 10 $\mu$m, then the motion of the photocenter due to the planet could be as high as several microarcseconds, and can often be of a much larger magnitude than seen at optical wavelengths when the photocenter motion is due solely to stellar motion.

We performed numerical modeling of several exoplanet systems via the {\sc reflux} code, and found it to be consistent with the predictions of our analytical model. The numerical modeling revealed that, even at shorter wavelengths where $\alpha$ $>$ 0, the planet has a visible impact on the observed astrometric orbit of the system. As well, deviations from pure sinusoidal motions due to day-night flux differences are clearly visible, and thus multi-wavelength astrometry could probe planetary properties of albedo and heat redistribution efficiency.

One caveat when working to extract the planetary and stellar masses from actual observations is that one will likely need to either precisely know the luminosity ratio of the system, or make assumptions about the luminosity of the planet, e.g., it radiates as a blackbody and is in thermal equilibrium. It may be possible that other observations could yield this information, such as the secondary eclipse depth if the planet happens to transit. The remaining parameters of the system's distance and period should be well determined via other methods such as microarcsecond precision parallax and radial-velocity or photometric light curves.

For the prospects of detection, it is clear that this effect will probably not be detected in the very near-term. Although astrometric measurements are approaching 1 $\mu$as accuracy, they have not yet been performed. Much of the ground-based work is being focused on the optical and K bands, where in the latter the effect is just barely detectable. The development of microarcsecond precision astrometric systems in the mid-infrared, or sub-microarcsecond precision in the near-infrared, are clearly needed, and the methods presented here will serve to preselect the best planetary system candidates to be observed by those systems.

The work presented in this paper assumed that both the star and planet radiate as blackbodies, however it is known that both can significantly deviate from that assumption, especially in the near infrared \citep[e.g.,][]{Gillon2009,Rogers2009,Gibson2010,Croll2011,deMooij2011,Coughlin2012}. At the extreme end, \citet{Swain2010} and \citet{Waldmann2012} recently found evidence for a very large non-LTE emission feature around 3.25 $\mu$m in the atmosphere of HD 189733 b\footnotemark[4]. Although via blackbody approximations we calculate that the planet-to-star flux ratio should be 8.3$\times$10$^{-4}$, \citet{Swain2010} and \citet{Waldmann2012} measure the 3.25 $\mu$m emission feature to be $\sim$8.5$\times$10$^{-3}$ times the stellar flux, or about ten times greater than expected. Assuming blackbody emission, the expected value for $\alpha$ for this system at visible wavelengths is 2.15 $\mu$as, and at 3.25 $\mu$m is 0.83 $\mu$as. If the emission feature is real however, the expected value for $\alpha$ at 3.25 $\mu$m is a very large -11.3 $\mu$as, dominated due to the planetary motion. Thus, the key in performing these types of observations may be to select particular wavelengths where the planets are unusually bright.

\footnotetext[4]{We note that \citet{Mandell2011} reported a non-detection of a portion of this feature between the publications of \citet{Swain2010} and \citet{Waldmann2012}.}

Finally, although we did not assume any limb-darkening in our models since we were examining near to mid-infrared wavelengths, limb-darkening will be significant when observed at different optical bandpasses. The astrometric signature of transiting planets will vary greatly due to limb-darkening in the optical regime, and thus multi-wavelength astrometry of transiting planets may be used to explore the limb-darkening profiles of stars, or visa versa, stellar limb-darkening may need to be precisely understood in order to extract planetary and stellar parameters of interest.

\acknowledgments
J.L.C acknowledges support from a NSF Graduate Research Fellowship. We thank Dawn Gelino and Thomas Harrison for heading up the SIM Science Study which started these series of papers. This research has made use of the Exoplanet Orbit Database and the Exoplanet Data Explorer at exoplanets.org. This research has made use of NASA's Astrophysics Data System.

\bibliography{AstroRefs.bib}

\end{document}